\documentclass{article}%
\usepackage{eurosym}
\usepackage[top=3cm, bottom=3cm, left=3cm, right=3cm]{geometry}
\usepackage{amsmath}
\usepackage{amsfonts}
\usepackage{amssymb}
\usepackage{graphicx}%
\usepackage{epsfig}
\begin{document}

\title{Study of Electromagnetic Properties of Light Baryons in the Hypercentral Approach}
\author{Zahra Ghalenovi \\Department of Physics, Kosar University of Bojnord, Iran
}
\maketitle
\begin{abstract}
Light flavour baryons are studied in a non-relativistic potential model with colour Coulomb plus linear confinement potential using a simple variational method. The ground-state masses and magnetic moments of the light baryons are computed using a spin- and isospin-dependent potential. We extend our scheme to predict the transition magnetic moments of $B_{(J^P={3/2}^+)} \rightarrow B_{(J^P={1/2}^+)}\gamma$ processes. We also compute the radiative decay widths and branching ratios of light baryons. A comparison of our results with those of other works and experimental data is also presented.\\


\end{abstract}

\section{ Introduction}
The baryon structures are recently studied by various constituent quark models \cite{1,2,3,4,5,6,7,8,9,10}. Theoretically, there exist serious discrepancies between the quark model and experimental data particularly in the predictions of baryon magnetic moments. Hadrons derive their magnetic moments from the quark-gluon dynamics of their underlying structure. However, due to the complexity of low-energy quantum chromodynamics (QCD), a more detailed understanding of the hadron magnetic moments seems difficult\cite{sharma}. After the advent of QCD, our knowledge of hadron’s magnetic moment calculations comes mostly from models: non-relativistic quark models, bag models, the Skyrme model,string models, the Nambu-Jona-Lasinio model. There now exist measurements of magnetic moments of all the octet baryons, except for the $\Sigma^{0}$ 
which has a life time too short for it to travel a significant distance even at the higher
energies now available \cite{Amsler}. For the decuplet baryons with $J^{P}=3/2^{+}$,  the experimental measurements are poor as they have very short life times due to available strong interaction decay channels \cite{Thakkar}. Non-relativistic quark models have proved very successful in describing hadronic properties \cite{Vijaya Kumar, Semay, Blask, Hassanabadi:2008zz}. In these models the hamiltonian usually contains three main ingredients: the kenetic energy, the confinement potential and a hyperfine interaction term, which has often been taken as an effective one-gluon-exchange potential (OGEP).  The color hyperfine splitting of OGEP is defined as the mass difference between two similar baryons that differ by one unit of spin (e.g. $\Delta_{B}=M_{B^{*}}-M_{B}$). In the present work we extend the previous models and also consider the flavour hyperfine splitting (taken from the one-Goldstone boson-exchange potential), which depends on the isospin of the baryons. Including the spin-isospin and isospin-isospin interactions to the hamiltonian makes us to obtain the better results for the baryons properties.\\
The purpose of this paper is to study the baryonic system using a simple variational method. We calculate masses and magnetic of ground-state light baryons using a potential model in the non-relativistic limit. We calculate the baryon spectrum in a two-step procedure: first, we use the hypercentral model in order to solve the three-body Schr\"{o}dinger equation performing a variational method and obtain the wavefunction and eigenvalues of the barionic system. Second, we consider the spin-spin, spin-isospin and isospin-isospin dependent potentials as perturbating hyperfine interaction and obtain the baryon masses. This paper is organized as follows. In Sec.\ 2 we introduce our potential model. We explain the hypercentral model and solve the Schr\"{o}dinger equation using the variational approach in Sec. \ 3. Our predictions for masses, magnetic moments are presented in Sec. 4. Applying the obtained results we calculate the transition magnetic moments, decay widths and branching ratios of $B_{3/2}\rightarrow B_{1/2}\gamma$ processes in Sec. 5, and  Sec.\ 6 includes conclusions.\\

\section{Interaction Potential Model}
To make a phenomenological model, we should introduce a potential model such that the QCD concepts of the quark-quark interactions be satisfied. From the experimental observations, we find that all hadrons are made of quarks and no single quark can be individually observed. This fact imply that the quarks are confined in hadrons in the vacuum. According to quantum chromodynamics, there are super-strong color attractive interactions among the quarks, causing three quarks of different colors to be confined together and form a colorless baryon. Moreover the experimental baryon spectroscopy shows an underlying SU(6) symmetry. Therefore an interaction potential should contain two main terms: a confining SU(6) invariant term, and a SU(6) breaking term describing the splittings of multiplets of baryons. Thus the three-quark interaction can be generally written in the form of

 \begin{equation}
V_{3q}=V_{SU(6)-invariant}+V_{SU(6)-breaking}.
\end{equation}

for the $V_{SU(6)-invariant}$ sector, the coulomb-plus-linear potential ($-\frac{a}{x}+bx$), 
known as the Cornell potential, has received a great deal of attention both in particle more precisely in the context of meson spectroscopy where it is describe systems of quark and antiquark bound states, and in atomic molecular physics where it represents a radial Stark effect in hydrogen. This potential model is a combination of lattice QCD calculations plus the Isgur-Karl interaction \cite{Karl 1, Karl 2} and includes the short distance Coulombic interaction of quarks and the large distance quark confinement, known from lattice QCD, via the linear term in a simple form. Coulombic term alone is not sufficient because it would allow free quarks to ionize from the system. The nonperturbative SU(6)-invariant part is introduced as follows:

 \begin{equation}  \label{confining}
V_{SU(6)-invariant}=\Sigma_{i<j}^{3}(ar_{ij}-\frac{c}{r_{ij}}),
\end{equation}

where $r_{ij}$ is the relative distance between $i$th and $j$th quarks and $a$ and $c$ are constant.\\
The SU(6)-breaking interaction is considered as perturbative term:

\begin{equation} \label{non-confining}
V_{SU(6)-breaking}=V_{Spin}+V_{Isospin}+V_{Spin-Isospin}.
\end{equation}

 In the $SU(3)_F$ invariant approximation the pseudoscalar exchange interaction splits the multiplets of $SU(6)_{SF}\times U(6)_{conf}$ in the spectrum to multiplets of  $SU(3)_F\times SU(2)_{s}\times U(6)_{conf}$ . The position of these multiplets differ in the baryon sectors with different strangeness because of the mass splitting of the pseudoscalar octet and the different constituent masses of the u, d and s quarks that breaks $SU(3)_F$ flavor symmetry. The simple representations of  one gluon exchange (OGE) and Goldstone boson exchange (GBE) interactions in the $SU(3)_F$ invariant limit are 

\begin{equation}  \label{OGE}
V_{OGE}\approx-\sum_{i<j}V(\vec{r}_{ij})\vec{\lambda}_{i}^{c}.\vec{\lambda}_{j}^{c}\vec{\sigma}_{i}.\vec{\sigma}_{j}
\end{equation}

 and
 
 \begin{equation}  \label{GBE}
V_{GBE}\approx-\sum_{i<j}V(\vec{r}_{ij})\vec{\lambda}_{i}^{F}.\vec{\lambda}_{j}^{F}\vec{\sigma}_{i}.\vec{\sigma}_{j},
\end{equation}

respectively, where the $\lambda^{c}$ $^{,}s$ and $\lambda^{F}$ $^{,}s$  are color Gell-Mann and SU(3) flavour  matrices  and the $i$ and $j$ sums run over the constituent quarks. The radial part $V(r)$ behaves as the usual Yukawa behaviour at long range, but at short range behaves as a smeared version of the $\delta$ function which needs to be regularized for practical calculations. It is the latter which plays a major role in describing the baryon spectra in the frame of Goldstone boson exchange models \cite{Stassart, Ghalenovi 2013}. Therefore, we consider only the delta function as the radial part of $V(r)$:

\begin{equation}\label{delta function}
V(r)\sim\delta(r)\approx\frac{1}{(\sqrt{\pi}\sigma_i)^3}exp(-\frac{r^2}{\sigma_{i}^{2}}),
\end{equation}

where $\sigma_i$ is constant and has different values for the GBE and OGE interactions. Therefore the non-confining interaction Eq. \ref{OGE} (or spin-spin interaction) contains a $\delta$ -like term that we
modify it by a Gaussian function 

\begin{equation}   \label{Hs}
H_S=\Sigma_{i<j}A_{S}(\frac{1}{\sqrt{\pi}\sigma_{S}})^3exp(\frac{-r_{ij}^2}{\sigma_S})(\overrightarrow{S_{i}}.\overrightarrow{S_j}),
\end{equation}

where $\overrightarrow{S_i}$ is the spin operator of the $i$ th
quark. The non-confining potential (\ref{GBE}) is provided by the
 Goldstone boson exchange  interactions, which gives rise to a 
 spin-and isospin-dependent part \cite{Golzman1, Golzman2}. \\
Recently, it has also been pointed out that an isospin dependence of 
the quark potential can be obtained by means of quark exchange (or GBE). 
More generally, one can expect that the quark-quark pair production 
can lead to an effective quark interaction containing an isospin (or
flavor)-dependent term \cite{Geiger}. With these motivations in mind,
we have introduced isospin-dependent terms. Finally, we add
two terms in the Hamiltonian quark-quark pairs with hyperfine 
interaction similar to Eq.(\ref{Hs}).  The first one depends on the
isospin only and has the form \cite{Geiger, Giannini 1,Giannini 3, Ghalenovi:EPJP, Ghalenovi:2014swa}

\begin{equation}  \label{Hi}
H_I=\Sigma_{i<j}A_{I}(\frac{1}{\sqrt{\pi}\sigma_{I}})^3exp(\frac{-r_{ij}^2}{\sigma_I^{2}})(\overrightarrow{t_i}.\overrightarrow{t_j}),
\end{equation}

where $\overrightarrow{t_i}$ is the isospin operator of the $i$ th
quarks. The second one is a spin-isospin interaction, given by
\begin{equation}  \label{Hsi}
H_{SI}=\Sigma_{i<j}A_{SI}(\frac{1}{\sqrt{\pi}\sigma_{SI}})^3exp(\frac{-r_{ij}^2}{\sigma_{SI}^{2}})(\overrightarrow{S_i}.\overrightarrow{S_j})(\overrightarrow{t_i}.\overrightarrow{t_j}),  
\end{equation}
where $\overrightarrow{S_i}$ and $\overrightarrow{t_i}$ are the spin
and isospin operators of the $i$ th quark respectively. Then from
Eqs. (\ref{Hs}-\ref{Hsi}), the hyperfine interaction is given by
\begin{equation}
H_{int}=H_S+H_I+H_{SI}.
\end{equation}
The contributions of this hyperfine interaction is added to the unperturbed SU(6)-invariant energies provided by the potential \ref{confining}. In the next section, we obtain the wave function and energy of system in the 
framework of a simple approximation with confining potential \ref{confining}.\\

\section{The Hypercentral model}

To describe the baryon as a bound state of three constituent
quarks, we define the configuration of three particles by the
Jacobi coordinates $\rho$ and $\lambda$ as 
\begin{equation}  \label{rhoo}%
\vec{\rho} = \frac{1}{\sqrt{2}}(\vec{r_1} - \vec{r_2}), \qquad
\vec{\lambda} = \frac{1}{\sqrt{6}}(\vec{r_1} + \vec{r_2} - 2
\vec{r_3})  ,
\end{equation}
such that
\begin{equation}
m_{\rho} = \frac{2 m_1 m_2}{m_1 + m_2}, \qquad m_{\lambda} =
\frac{3 m_3 (m_1 + m_2)}{2 (m_1 + m_2 + m_3)} .
\end{equation}
Here $m_1$, $m_2$ and $m_3$ are the constituent quark masses.
Instead of $\rho$ and $\lambda$ , one can introduce the
hyperspherical coordinates, which are given by the angles
$\Omega_{\rho}=(\theta_{\rho},\phi_{\rho})$  with the
hyperradius $x$, and the hyperangle $\zeta$, defined respectively
by 
\begin{equation}
x = \sqrt{\rho^2 + \lambda^2}, \qquad \xi = \arctan
(\sqrt{\frac{\rho}{\lambda}}).
\end{equation}

Note that the hyperradius $x$ can be expressed as the average relative
coordinate of three quark pairs in the baryon, i.e.,
\begin{equation} \label{10}
x=\sqrt{\frac{(\vec{r}_1-\vec{r}_2)^2 + (\vec{r}_2-\vec{r}_3)^2 + (\vec{r}_3-\vec{r}_1)^2}{3}}.
\end{equation}

Therefore the Hamiltonian will be
\begin{equation}
H=\frac{p^{2}_{\rho}}{2m}+\frac{p^{2}_{\lambda}}{2m}+V(x).
\end{equation}

In hyperspherical coordinates the Laplace operator for three-body system is written as follows:
\begin{equation}    \label{laplace}
\nabla^{2}=(\nabla^{2}_{\rho}+\nabla^{2}_{\lambda})=-(\frac{d^{2}}{dx^{2}}+\frac{5}{x}\frac{d}{dx}-\frac{L^{2}(\Omega_{\rho},\Omega_{\lambda},\xi)}{x^{2}}),
\end{equation}

Therefore the kinetic energy operator of a three-body problem can be written as ($\hbar=c=1$):
\begin{equation}
-\frac{1}{2m}(\nabla^{2}_{\rho}+\nabla^{2}_{\lambda})=-\frac{1}{2m}(\frac{d^{2}}{dx^{2}}+\frac{5}{x}\frac{d}{dx}-\frac{L^{2}(\Omega_{\rho},\Omega_{\lambda},\xi)}{x^{2}}).
\end{equation}

The eigenfuctions of $L^{2}$ are hyperspherical harmonics
\begin{equation}
L^{2}(\Omega_{\rho},\Omega_{\lambda},\xi)Y_{[\gamma],l_{\rho},l_{\lambda}}(\Omega_{\rho},\Omega_{\lambda},\xi)=\gamma(\gamma+1)Y_{[\gamma],l_{\rho},l_{\lambda}}(\Omega_{\rho},\Omega_{\lambda},\xi).
\end{equation}

 $\gamma$ is the grand angular quantum number given by $\gamma=2n+l_\rho+l_\lambda$  ; $l_\rho$ and $l_\lambda$  are the angular momenta associated with the  $\rho$ and $\lambda$ variables and $n$ is a non-negative integer number. The wave function of any system containing three particles can be expanded in the hyperspherical harmonic basis as follows

\begin{equation}   \label{Psi}
\Psi(\rho,\lambda)=\Sigma_{\gamma,l_{\rho},l_{\lambda}}  N_{\gamma}\psi_{\gamma}(x)Y_{[\gamma],l_{\rho},l_{\lambda}}(\Omega_{\rho},\Omega_{\lambda},\xi).  
\end{equation}

In our previous studies, we have presented the different methods to solve the non-relativistic two- and three-body problems \cite{Ghalenovi:EPJP, Ghalenovi:2014swa, Ghalenovi:IJMPE, Ghalenovi:2016}. 

\subsection{The eigenfunctions and eigenenergies of the baryonic systems}
One can write the total wave function of the three-body baryonic system as a product of the spacial, spin, flavour and color wave functions:

\begin{equation}
\Psi=\psi_{space}\chi_{spin}\Theta_{isospin}\theta_{color}=\psi_{space}\Phi_{SU(6)}\Theta_{color}
\end{equation}

The color factor gives no contribution to the matrix elements of the observable quantities and will be always omitted as customary in the quark models \cite{8}.\\

In the hypercentral model the  Schr\"{o}dinger equation has the form of \cite{Ghalenovi:2016, Ghalenovi:2011zz, Ghalenovi:2011, Salehi4}

\begin{equation}\label{Schro}
[\frac{d^2}{dx^2}+\frac{5}{x}\frac{d}{dx}-\frac{\gamma(\gamma+4)}{x^2}]\Psi_{\gamma(x)}=-2m[E_\gamma-V]\Psi_{\gamma(x)},
\end{equation}

If the energy of the system $ V $ depends only on the quarks distance, the remaining hyperradial part of the wave function is determined by 

\begin{equation}\label{Schro 2}
[\frac{d^2}{dx^2}+\frac{5}{x}\frac{d}{dx}-\frac{\gamma(\gamma+4)}{x^2}]\psi_{\gamma(x)}=-2m[E_\gamma-V(x)]\psi_{\gamma(x)}.
\end{equation}

The interaction potential $ V$ (Eq. \ref{confining}) in terms of the hyperradius $x$ has them form

 \begin{equation} \label{V(x)}
V(x)=ax-\frac{c}{x} .
\end{equation}

In order to solve Eq. \ref{Schro 2} for the ground-state baryons (i.e. $\gamma=0$), we assume the transformation

\begin{equation}  \label{chi}
\psi(x)=x^{-5/2}\chi(x),
\end{equation}

then Eq. \ref{Schro 2} reduces to the form

\begin{equation} \label{Schro 3}
\frac{d^{2}\chi(x)}{dx^{2}}+2m[E-V(x)-\frac{15}{8m
x^{2}}]\chi(x)=0
\end{equation}

and $V(x)$ is the three-quark potentials over the six-dimensional
sphere that was defined in Eq. (1). 

We solve Eq. \ref{Schro 3} by the variational method. We introduce a
simple variational ansatz for $\chi(x)$ as
\begin{equation}
\chi(x)=2\sqrt{2}p^{3} x^{\frac{5}{2}}e^{-P^{2}x^{2}}
\end{equation}
where $p$ is the variational parameter, and the numerical factor
is chosen so that $\int\chi^{2}(x)dx=1$ . The trial three-quark
Hamiltonian admits explicit solutions for the wave function and
the energy $E_0=minE(p)$ where
\begin{equation}
E(p)=<\chi|H|\chi>
\end{equation}
Now by using the condition $\frac{dE}{dp}|_{p=p_0}=0$, the value
of $p_0$ is found.

\section{Masses and magnetic moments of Baryons}

\noindent Baryon mass is obtained by sum of the quark masses and the hyperfine interaction potential 
treated as a perturbation:

\begin{equation}\label{Mass}
M_{baryon}=\sum_{i=1}^{3}m_i+E_{0}+<H_{int}>,
\end{equation}

where using the unperturbed wave function \ref{chi} we have

\begin{equation}
<H_{int}>=\int\psi H_{int}\psi dx.
\end{equation}

$E_{0}$ is the first order energy correction from the nonconfining potential and depends on $a$ and $c$ parameters which are taken from Ref. \cite{Ghalenovi:2011zz} (listed in table \ref{tab:parameters}). For the present calculations, we employ the same mass parameters of the light flavour quarks (m$u$=m$d$ = 330 MeV, m$s$ = 500 MeV) as used in \cite{Ebert2, Ebert3, Lang Yu2}.   The hyperfine potential parameters are obtained by global fit to the experimental masses and magnetic moments of light baryons (see table \ref{tab:parameters}).\\

 \begin{table} [ptb] 
 \caption{The parameters of the potential model.  \label{tab:parameters}}
 \begin{center}
\begin{tabular}[c]{c|c}     \hline            
$a$  & 1.61 $fm^{-1}$\\
$c$ & 4.59 \\
$A_s$&  87.1   $(fm)^2$\\
$\sigma_S$& 2.31   $fm$\\
$A_I$ &  49.9   $(fm)^2$\\
$\sigma_I$ &  2.4   $fm$\\
$A_{SI}$ & 47.4   $(fm)^2$\\
$\sigma_{SI}$&  2.9  $fm$ \\

\end{tabular}
  \end{center}
   \end{table}

Within the baryons the mass of the quarks may get modified due to its binding interactions with the other quarks. The effective quark mass is defined as 

\begin{equation}\label{Meff}
m_{i}^{eff}=m_{i}(1+\frac{E_{\gamma}+<H_{int}>}{\sum_{i}m_{i}}),
\end{equation}

such that the mass of the baryon is

\begin{equation}\label{Mass2}
M_{baryon}=\sum_{i}m_{i}^{eff}.
\end{equation}

Therefore a special quark has different effective masses inside different baryons. The evaluated effective quark masses are listed in table \ref{tab:effectiv m}.  In table \ref{tab:mass} the obtained baryon masses are listed and compared with other theoretical predictions \cite{Ghalenovi:2012, Thakkar2,Giannini 4} or experimental data \cite{PDG1}. Refs. \cite{Ghalenovi:2012, Thakkar2,Giannini 4} calculated light baryon spectrum in the different non-relativistic hypercentral quark models. In the present work, we extend their calculations and also compute the electromagnetic properties of light baryons.\\

 \begin{table} [ptb] 
 \caption{Effective quark masses (in MeV) inside the different baryons. \label{tab:effectiv m}}
 \begin{center}
\begin{tabular}[c]{c|c|c}     \hline            
Baryon &  $m_u^{eff}=m_d^{eff}$ & $m_s^{eff}$\\ \hline  
$N$  &    313.1 & \\ 
$\Delta $ & 411.6 &  \\
$\Sigma$ & 328 & 496.9   \\
$\Lambda $&  311.5 & 472   \\
$\Sigma^*$&  383.2 & 580.6   \\
$\Xi $&  321 & 486.5   \\
$\Xi^*$ & 369.3 & 560   \\  
$ \Omega$  &   &  552.7 \\  \hline

\end{tabular}
  \end{center}
   \end{table}

The magnetic moment of baryon is obtained in terms of the spin-flavour wave function of the constituent quarks as

\begin{equation}\label{Mu}
\mu_{B}=\sum_{i}\left\langle \phi_{sf}|\mu_{i}\vec{\sigma}_{i}|\phi_{sf}\right\rangle,
\end{equation}

where 

\begin{equation}\label{mu}
\mu_{i}=\frac{e_i}{2m_{i}^{eff}}.
\end{equation}

Here, $e_i$ and $s_i$ represent the charge and the spin of the quark $(s_{i}=\frac{\sigma_i}{2})$ constituting the baryonic state and $|\phi_{sf}>$ represents the spin-flavour wave function of the respective baryonic state. The calculated magnetic moments of light baryons are presented in table \ref{tab:Magnet} and compared with the experimental data  \cite{PDG2, Kotulla, Bosshard} and other theoretical results \cite{Thakkar2,Bos, Dhir1}.

\section{Radiative decay width and branching ration}
The study of electromagnetic transitions of decuplet to octet baryons is an important issue for understanding of internal structure of baryons \cite{Negash, Ramollah, Oset, Lang Yu}. The prediction of electromagnetic properties of the both octet and decuplet baryons was not successful in the most non-relativistic phenomenological models \cite{Lang Yu, Isgur 1,Isgur 2, Isgur 3, Isgur 4}. In this work, we study the properties of the both octet and decuplet baryons.\\
The radiative decay width of the baryons is given by \cite{Majethiya 1}

\begin{equation}\label{Gama}
\Gamma_R=\frac{q^{3}}{4\pi}\frac{2}{2J+1}\frac{e^{2}}{m_{p}^{2}}|\mu_{\frac{3}{2}^{+}\rightarrow\frac{1}{2}^{+}}|^2,
\end{equation}

where $m_p$ is the proton mass, $ \mu_{\frac{3}{2}^{+}\rightarrow\frac{1}{2}^{+}} $ is the radiative transition magnetic moments and $ q $ is the photon energy. We obtain the transition moments by sandwiching Eq. \ref{chi} between the appropriate $3/2^{+}$ and $1/2^{+}$ baryon wave functions. The transition magnetic moments for $\frac{3}{2}^{+}\rightarrow\frac{1}{2}^{+}$ are computed as

\begin{equation}\label{Mu transition}
\mu_{\frac{3}{2}^{+}\rightarrow\frac{1}{2}^{+}}=\Sigma_{i} <\phi_{sf}^{\frac{3}{2}^{+}}|\mu_{i}\sigma_i|\phi_{sf}^{\frac{1}{2}^{+}}>.
\end{equation}

$<\phi_{sf}^{\frac{3}{2}^{+}}|$ represents the spin flavour wave function of the quark composition for the respective decuplet baryons while  $<\phi_{sf}^{\frac{1}{2}^{+}}|$  represents the spin flavour wave function of the quark composition for the octet baryons. The value of $\mu_i$ is given by Eq. \ref{mu}. Since the effective masses of the constituent quarks are different in the different baryons therefore, in order to evaluate $B_{3/2}\rightarrow B_{1/2}\gamma$  transition magnetic moments, we take the geometric mean of effective quark masses of the constituent quarks of initial- and final-state baryons:
\begin{equation}
m^{eff}=\sqrt{m^{eff}_{i(\frac{3}{2}^{+})}m^{eff}_{i(\frac{1}{2}^{+})}}  ,
\end{equation}

where $m^{eff}_{i}$ is the effective quark mass of the $i$th quark inside the corresponding baryon. Our results for the transition magnetic moments and decay widths are listed in tables \ref{tab:magnetic tr} and  \ref{tab:Gamma} respectively and, compared with the other results \cite{Lang Yu2, aliev 1} and experimental data \cite{Bosshard, Eidelman, Colas, Mast, Molchanov}. The authors of Ref. \cite{Lang Yu2} studied the transition magentic moments and decay widths of light baryons employing a chiral quark model based on the Goldstone boson exchanges. In Ref. \cite{aliev 1} the electromagnetic properties of the light baryons in the framework of the light cone QCD sum rules are computed. In comparison with the models presented in Res. \cite{Lang Yu2, aliev 1}, this work presents a simple model in which addition to the barionic decay widths, we get also a good baryon spectrum.
On the other hand, the results listed in table \ref{tab:Gamma} show that  in this model, the computed decay widths of $ \Delta $ baryons are closer to the experiments. 
Using the experimental total decay width $\Gamma(Baryon)$ of the respective decuplet baryons, we calculate the branching ratios $\frac{\Gamma_R}{\Gamma(Baryon)}$ and compare them with the experiment \cite{Amsler} (see table \ref{tab:ratio}).\\

\section{Conclusions}
In this paper the properties of the light baryons are studied in the hypercentral quark model. We solved the Schr\"{o}dinger equation in a variational approach and found the eigen-energies and eigen-functions of the baryons. Using the theory of time-independent perturbation for the hyperfine interactions, we got the effects of spin and isospin potentials in the shift of baryon energy. The computed errors in the last columns of table \ref{tab:mass} and also the calculated magnetic moments listed in table \ref{tab:Magnet} show that our model is successful in describing the baryon properties.  Introducing the effective quark masses, we calculated  $B_{3/2}\rightarrow B_{1/2}\gamma$  transition magnetic moments and radiative decay widths. The computed $\Delta \rightarrow n \gamma$ transition magnetic moment is lower than experimental value 3.23 \cite{Bosshard}. This discrepancy is also found in other theoretical models \cite{Dhir1, Lang Yu2, Hong}. But the results for the radiative decay widths and branching ratios are in very good agreement with experimental data.\\

\begin{table}[ptb] 
    \caption{ Mass spectrum of the light baryons (in MeV). The percentage of relative errors between our results and experimental data are represented in the last column.  \label{tab:mass}}

    \centering
    \begin{center}
\begin{tabular} [c]{c|c|c|c|c|c|c|c} \hline 
Baryon&Exp\cite{PDG1}&Present work&\cite{Ghalenovi:2012}&\cite{Thakkar2}& \cite{Giannini 4}  & Error\\  \hline 
N(938)P11 & 938           & 939 &     938&     931&     938   & $0.1\%$\\
$\Delta$(1232)P33&  1231-1233   & 1234& 1220&    1228   & 1232  & $0.0\%$\\
$\Lambda(1116)P_{01}$    & 1116 & 1095 & 1133 & 1147 & 1116  & $1.8\%$  \\ 
$\Sigma(1193)P_{11}$    & 1193 & 1153 & 1170 & 1198 & 1193 &  $3.3\%$ \\ 
$\Sigma^*(1385)P_{13}$    & 1383-1385 & 1347 & 1382 & 1392 & 1371&  $2.6\%$ \\ 
$\Xi(1318)P_{11}$       &1314-1316 & 1294 & 1334 & 1297   & 1332  &  $1.8\%$ \\
$\Xi^*(1530)P_{13}$       &1531-1532 & 1490 & 1524 &   1516 & 1511  & $2.6\%$  \\
$\Omega(1672)P_{03}$ &  1672-1673 &   1658   & 1670 &     1678&    1650.5   & $0.8\%$\\ \hline

\end{tabular}
\end{center}
\end{table}

 \begin{table} [ptb] 
\begin{center}
\caption{Magnetic moments of light baryons in terms of nuclear magneton $\mu_N$.   \label{tab:Magnet}}
\begin{tabular}[c]{c|c|c|c|c|c|c|c} \hline
Baryon&Present work&      Exp \cite{PDG2, Kotulla, Bosshard}& \cite{Thakkar2} &\cite{Bos}& \cite{Dhir1}  \\  \hline
$p$            & 2.99    & 2.79                & 3.04  & 2.79  &       \\
$n$            & -1.99   & -1.91               & -2.07  & -1.97  &       \\
$\Delta^{++}$  & 4.55    & $4.5\pm 0.95$       & 4.66  &   &   4.56   \\
$\Delta^{+}$   & 2.27    & $2.7^{+1.0}_{-1.3}$ &  2.35 &   &   2.28   \\
$\Delta^{0}$   &$ \sim $0& $ \sim $0           &  0.05 &   &   0.0    \\
$\Delta^{-}$   & -2.27   &                     &  -2.33 &   &   -2.28    \\ 
$\Lambda$       & -0.66 & -0.61 & -0.64 & -0.60 &  \\
$\Sigma^{+}$    & 2.75 & 2.46 & 2.63 & 2.48 &  \\
$\Sigma^{*+}$    & 2.72 &   & 2.60 &  & 2.56 \\
$\Sigma^{0}$    & 0.84&  & 0.83 &  0.66 &   \\
$\Sigma^{*0}$    & 0.27&  & 0.28 &  &  0.23 \\
$\Sigma^{-}$    & -1.06 & -1.16 & -0.98 & -1.16 &   \\
$\Sigma^{*-}$    & -2.17 &  & -2.40 &  & -2.10  \\
$\Xi^{0}$       & -1.50 & -1.25 & -1.49 & -1.27 &  \\ 
$\Xi^{*0}$       & 0.57 &   & 0.53 &  & 0.48 \\
$\Xi^{-}$       & -0.53 & -0.65 & -0.55 & -0.59 &  \\
$\Xi^{*-}$       & -1.96 & & -1.91  &  & -1.90 \\
$\Omega$       & -1.70 & -2.02$\pm 0.06$ & -1.67 &  & -1.67 \\ \hline

\end{tabular}
\end{center}
\end{table}

\begin{table}[ptb]  
    \caption{ Magnitude of the transition magnetic moments $(|\mu_{\frac{3}{2}^{+}\rightarrow\frac{1}{2}^{+}}|)$ in $\mu_{N}$.            \label{tab:magnetic tr}}
    \centering
    \begin{center}
    \begin{tabular} [c]{c|c|c|c|c}
    \hline
     Decay Mode& Present work&   Exp \cite{Bosshard}  &\cite{aliev 1}  & \cite{Lang Yu2} \\
\hline
$\Delta^{+}\rightarrow p\gamma$  &2.47 & $3.23\pm 0.1$  & 2.50    & 2.57 \\
$\Delta^{0}\rightarrow n\gamma$& 2.47 & $3.23\pm 0.1 $    &2.50   & 2.57 \\
$\Sigma^{*+}\rightarrow\Sigma^{+}\gamma$ & 2.21&  &   2.10 & 2.21  \\
$\Sigma^{*0}\rightarrow\Sigma^{0}\gamma$& 0.83 &  & 0.89  & 0.88 \\
$\Sigma^{*0}\rightarrow\Lambda^{0}\gamma$&2.21&  & 2.30 & 2.24 \\
$\Sigma^{*-}\rightarrow\Sigma^{-}\gamma$&0.28&   & 0.31  & 0.44 \\
$\Xi^{*0}\rightarrow\Xi^{0}\gamma$& 2.27  &       & 2.20  & 2.22\\
$\Xi^{*-}\rightarrow\Xi^{-}\gamma$& 0.29 &  & 0.31  &   0.44 \\
\hline
    \end{tabular}
    \end{center}
\end{table}

\begin{table}[ptb] 
    \caption{ Radiative decay widths $\Gamma_{R}$ (in MeV).   \label{tab:Gamma}}
    \centering
    \begin{center}
    \begin{tabular} [c]{c|c|c|c|c|c}
    \hline
     Decay Mode    &Present work&   Exp \cite{Eidelman, Colas, Mast, Molchanov}&  \cite{aliev 1} & \cite{Lang Yu2} \\
\hline
$\Delta^{+}\rightarrow p\gamma$& 0.648 &0.64   &   0.90 &  0.363  \\
$\Delta^{0}\rightarrow n\gamma$ & 0.648  &0.64  &0.90   & 0.363   \\
$\Sigma^{*+}\rightarrow\Sigma^{+}\gamma$ & 0.149&  & 0.11  & 0.100  \\
$\Sigma^{*0}\rightarrow\Sigma^{0}\gamma$& 0.021 &$ <1.750$ & 0.021  & 0.016  \\
$\Sigma^{*0}\rightarrow\Lambda^{0}\gamma$&0.325& $<2.100$   & 0.470 & 0.241\\
$\Sigma^{*-}\rightarrow\Sigma^{-}\gamma$&0.002& $<0.009$    & 0.002 &0.004 \\
$\Xi^{*0}\rightarrow\Xi^{0}\gamma$& 0.158  &   & 0.140 & 0.131\\
$\Xi^{*-}\rightarrow\Xi^{-}\gamma$& 0.002 &  &  0.003  &  0.005  \\
\hline 
    \end{tabular}
    \end{center}
\end{table}

\begin{table}[ptb] 
    \caption{Branching ratios $\frac{\Gamma_{R}}{\Gamma_{\Delta}}$ (in $ \%$).   \label{tab:ratio}}
    \centering
    \begin{center}
    \begin{tabular} [c]{c|c|c|c|c|c}
    \hline
     Decay Mode   &Present work&   Exp \cite{Amsler}  \\
\hline
$\Delta^{+}\rightarrow p\gamma$& 0.549 & 0.52-0.60   \\
$\Delta^{0}\rightarrow n\gamma$&  0.549  & 0.52-0.60    \\
$\Sigma^{*+}\rightarrow\Sigma^{0}\gamma$& 0.416 &    \\
$\Sigma^{*0}\rightarrow\Sigma^{0}\gamma$& 0.058 &    \\
$\Sigma^{*0}\rightarrow\Lambda^{0}\gamma$&  0.906  & $ 1.3\pm0.4 $    \\
$\Sigma^{*-}\rightarrow\Sigma^{-}\gamma$& 0.007 &$ <0.024 $  \\
$\Xi^{*0}\rightarrow\Xi^{0}\gamma$&  1.740 & $ <4.0 $    \\
$\Xi^{*-}\rightarrow\Xi^{-}\gamma$& 0.029 &$ <0.4 $  \\
\hline 

    \end{tabular}
    \end{center}
\end{table}

\end{document}